\documentstyle[12pt,epsf]{article}
\begin{document}
  \bibliographystyle{prsty}
  \baselineskip 16pt plus 2pt minus 2pt

  \newcommand{\beq}{\begin{equation}}
  \newcommand{\eeq}{\end{equation}}
  \newcommand{\beqa}{\begin{eqnarray}}
  \newcommand{\eeqa}{\end{eqnarray}}
  \newcommand{\dida}[1]{/ \!\!\! #1}
  \renewcommand{\Im}{\mbox{\sl{Im}}}
  \renewcommand{\Re}{\mbox{\sl{Re}}}
  \def\simge{\hspace*{0.2em}\raisebox{0.5ex}{$>$}
       \hspace{-0.8em}\raisebox{-0.3em}{$\sim$}\hspace*{0.2em}}
  \def\simle{\hspace*{0.2em}\raisebox{0.5ex}{$<$}
       \hspace{-0.8em}\raisebox{-0.3em}{$\sim$}\hspace*{0.2em}}
  
  \begin{titlepage}
  
  \hfill{TRI-PP-99-32}
  
  \vspace{1.0cm}
  
  \begin{center}
  {\large {\bf Astrophysical Implications of the Induced Neutrino Magnetic Moment 
from Large Extra Dimensions }}
  
  \vspace{1.2cm}
  
  G. C. McLaughlin\footnote{email: gail@lin04.triumf.ca} and
  J. N. Ng\footnote{email: misery@triumf.ca}
  
  \vspace{0.8cm}
  
  TRIUMF, 4004 Wesbrook Mall, Vancouver, BC, Canada V6T 2A3\\[0.4cm]
  \end{center}
  
  \vspace{1cm}
  
  \begin{abstract}
  
  Theories involving extra dimensions, a low ($\sim$ TeV) string scale
  and bulk singlet neutrinos will produce an effective neutrino magnetic moment
  which may be large ($\simle 10^{-11} \mu_B$).  The effective magnetic
  moment increases with neutrino energy, and therefore high energy reactions
  are most useful for limiting the allowed number of extra dimensions.   
  We examine constraints from both neutrino-electron scattering and also astrophysical
  environments.  We find that supernova energy loss considerations require 
  a number of extra dimensions, $n \ge 2$, for an electron neutrino-bulk neutrino
Yukawa coupling of order 1.    
  
  \end{abstract}
  
  \vspace{2cm}
  \vfill
  \end{titlepage}
  
  %\section{Introduction}
  
  The study of string theory has yielded many beautiful theoretical results. Although
  direct confrontation with experiments is still elusive, the effort has 
  nonetheless  given us many new ways of looking at the physical world.  Weak coupling
string theory predicts \cite{Gin}
  \beq
      M_{S}= M_{\mathrm P} {\sqrt {k\alpha_{GUT}}}, \ 
  \eeq
where  $\alpha_{GUT}$ is the unified gauge coupling constant and $k$ is an 
  integer of order one denoting the level of the Kac-Moody algebra.  In this
case the string scale, $M_S$ is close to the Plank scale, $M_P$.
Recently,  Witten \cite{Wit}
observed that in nonperturbative string theory, $M_S$ could
be lower.  Lykken \cite{Lyk} and Antoniadis \cite{ant} discussed the 
possibility that the string scale 
could be close to the weak scale, which has many interesting phenomenological 
consequences.  If the string scale where gravity becomes strong is around the
TeV scale, one needs to understand why the observed Newton's constant at large 
  distances is small. Here one elicits the help of extra dimensions inherent in string 
theories.
  It is well known that string theory can be consistently formulated in 10 or 11 dimensions. 
  Hence 6 or 7 extra spatial dimensions exist which must be compactified. 
  The weakness of gravity is then due its spreading  in 
  these $n$ extra dimensions \cite{Nima}. The relation
  that replaces Eq.(1) is given by
  \beq
  \label{eq:fun}
        M^2_{\mathrm P}= M^{n+2}_{\ast}V_n \;
  \eeq
where $V_n$ is the volume of the extra space and $M_{\ast}$ is the higher dimensional Planck 
scale.  The two scales $M_S$ and $M_{\ast}$ are related but the exact relation is not needed in this 
investigation.  For toroidal compactification we have 
  \beq
         V_n= (2\pi)^nR_1R_2\ldots R_n \; 
  \eeq
  and $R_i(i=1,2,\ldots n)$are the radii of the extra dimensions. 
  For simplicity usually one sets all the radii to be
   equal and denoted by $R$, which is referred to as the symmetric compactification case. 
  These theories predict a deviation from the Newtonian $1/r$ law of gravitational potential
  at small distances.  The fact that this law is
   well  tested above  a distance of 1 mm leads to the conclusion that
   $n \geq 2$ for $M_\ast = 1 \, \mathrm TeV$ for the symmetric case.  Current proposed experiments 
  may test smaller distance scales \cite{Lon}. Recently \cite {Rand} proposed an alternative to the
  usual compactification of extra dimensions.  Here,  the important effect of the
extra dimension is to alter the four dimensional metric term.  In this scenario 
the extra dimensions are not compactified and 
gravitational experiments are not necessarily a test of the fundamental 
scaling law of Eq.~\ref{eq:fun}.

Independently of the exact nature of the scaling law to which gravity is sensitive, 
theories which involve 
extra dimensions can also provide a mechanism for 
producing a small neutrino mass.  In the following, we briefly review this mechanism
and then point out that it will also generate a large
neutrino magnetic moment.  This can be used to put constraints on the number of extra
dimensions involved in producing the neutrino mass.
  
  It is well known that the Standard Model (SM) is phenomenally successful in describing experiments.
  To accommodate this in the above described extra dimensional world one localizes the SM fields on
a hypersurface with three spatial dimensions or 3-brane \cite{Pol}.
 Due to the conservation of gauge flux all 
  charged states under the SM group are expected to be trapped on the 3-brane \cite{Dva}. 
  Such a scenario with the SM residing on a 3-brane and gravity 
propagating in the bulk of a 
  10 or 11-dimensional world has thus far not been ruled out by experiments for 
  $n \geq 2$ and $ M_\ast \geq \mathrm O(10 \, TeV)$. To date the most 
  stringent constraint on this scenario comes from analysis of
  supernova energy loss due to Kaluza-Klein graviton emission 
  and sets $M_\ast \geq 30 \sim 50 \, \mathrm TeV$ \cite{Cul} for $n = 2$
  symmetric compactification\cite{many}.  While the SM fields are localized on the 3-brane, 
there is no argument to suggest that SM singlet fields need to do the same. Thus, any gauge 
singlet field is allowed to be bulk matter just like the graviton. 

Theories with extra dimensions provide an elegant new mechanism to generate naturally
small neutrino masses,  $m_\nu \leq 10^{-3} \mathrm eV$. Such small neutrino masses are indicated by
  recent experiments in solar and atmospheric neutrinos \cite{Supk}. The key here is to postulate the
existence of a right-handed neutrino, $\nu_R$, which is a bulk fermion and use it to generate a small 
  neutrino mass \cite{Nima2,Die,pos}, 
by coupling to the SM left-handed  neutrino field $\nu_L$ 
  via the standard Higgs mechanism with a crucial suppression factor. This arises because in 
  general the couplings between states on the brane and bulk states
  are suppressed by a volume factor of $M_\ast^{-\frac{n}{2}}V_n^{-\frac{1}{2}}$. 
If combined with the scaling law of Eq. \ref{eq:fun}, 
the neutrino-Higgs interaction is given by
  \beq
  \label{eq:hnu}
  \frac{y}{\sqrt{M_\ast^nV_n}}H\bar{\nu}_{L}\nu_R=\frac{yM_\ast}{M_{\mathrm P}}H\bar{\nu}_{L}\nu_R \; ,
  \eeq
  where $H$ denotes the Higgs doublet, $y$ is the Yukawa coupling and $\nu_R$ is taken to be 
  the bulk fermion. Spontaneous electroweak symmetry breaking then generates a Dirac mass with the active 
  neutrino mass given by
  \beq
  \label{eq:dm}
           m_{D} = {yv \over \sqrt{M_\ast^nV_n}} 
  \eeq
  where $v=247 \, \mathrm GeV$. If we assume that $y=1$, the
scaling relation of Eq.~\ref{eq:fun} and $M_{\ast}\sim \mathrm 10 \, TeV$ 
  then $m_{D}\sim \mathrm 10^{-4} \, eV$.  
  
  The state $\nu_R$ is a bulk state and hence is a linear combination of a tower of 
  Kaluza-Klein (KK) states.  In the following, we will assume that $\nu_R$ is a Dirac
particle.  In this case, there is a tower of left handed states $\nu_{kL}$.
We first illustrate the relationship of the Kaluza-Klein towers to the active neutrinos
$\nu_e$ with the case of one dimension, compactified on  a circle. To further simplify
the model we assume that the higher dimensional bare Dirac mass term vanishes. This
can be implemented naturally under ${\mathcal Z}_2$ orbifold compactification [for details of
this scheme see \cite{Die}]. 
The quantization of internal momenta in the extra dimension generates Dirac mass
terms (for details see \cite{Nima2,Die})  
  \beq
  \label{eq:msum}
              \sum_{k=-\infty}^{\infty}
   m_k\bar{\nu}_{kR}\nu_{kL} +h.c. , \hspace{2.cm}  m_k= \frac{k}{R}  \;
  \eeq
  with a mass splitting of $1/R$. The mixing of the bulk states with the active $\nu_{eL}$ 
  gives rise to universal Dirac mass terms 
  \beq
  \label{eq:muni}
           m_{D} \sum_{k=-\infty}^{\infty} \bar{\nu}_{kR}\nu_{eL} + h.c. \;
  \eeq
  With regard to the electroweak interactions $\nu_{kL}$ and $\nu_{kR}$
 are sterile states. Expanding Eqs. \ref{eq:msum} and \ref{eq:muni}  
  gives the mass terms as:
  \beq
  \label{eq:nud}
      m_{D}\bar{\nu}_{0R}\nu_{eL} + m_{D}\sum_{k=1}^{\infty}(\bar{\nu}_{kR}+\bar{\nu}_{-kR})\nu_{eL}+
  \sum_{k=1}^{\infty}\frac{k}{R}(\bar{\nu}_{kR}\nu_{kL}-\bar{\nu}_{-kR}\nu_{-kL}) + h.c.   \;
  \eeq
   
  It is useful to define the following orthogonal states
  \beq
  \nu^{'}_{kR} = \frac{1}{\sqrt {2}}(\nu_{kR} + \nu_{-kR}) \hspace{2.cm} 
  \nu^{''}_{kR} = \frac{1}{\sqrt{2}}(\nu_{kR}-\nu_{-kR})  \:
  \eeq
  and
  \beq
  \nu^{'}_{kL} = \frac{1}{\sqrt {2}}(\nu_{kL} - \nu_{-kL}) \hspace{2.cm} \nu^{''}_{kL} = \frac{1}{\sqrt{2}}(\nu_{kL} + \nu_{-kl}) \;,
  \eeq
which can be used to rewrite Eq.\ref{eq:nud} as
  \beq
  \label{eq:nudd}
        m_D\bar{\nu}_{0R}\nu_{eL}+\sqrt {2}m_D\sum_{k=1}^{\infty}\bar{\nu}^{'}_{kR}\nu_{eL}+\sum_{k=1}^{\infty}\frac{k}{R}(\bar{\nu}^{'}_{kR}\nu^{'}_{kL}+\bar{\nu}^{''}_{kR}\nu^{''}_{kL}) + h.c.  \:
  \eeq
  The states with double prime superscripts have no low energy interactions and will be 
  ignored in our discussions.
  The mass term can now be written in the familiar form of $\overline{\nu}_{L}M\nu_{R}$ where
  $\nu_{L}=(\nu_{eL},\nu_{1L}^\prime,\nu_{2L}^\prime,\ldots)$ and 
$\nu_{R}=(\nu_{0R}^\prime,\nu_{1R}^\prime,\nu_{2R}^\prime,\ldots)$. The mass matrix
  M for $k+1$ states look as follows
  
  \beqa
  M=\left[ \begin{array}{ccccc}
  m_{D}&\sqrt {2}m_{D} &\sqrt {2}m_{D}  &\dots  &\sqrt {2}m_{D} \\
  	0    &\frac{1}{R}  &0           &\dots&  0            \\
  	0    &   0         &\frac{2}{R} &\dots&  0             \\
  	\dots& \dots       & \dots       &\dots&  \dots         \\
  	0    &  0          &  0          &\dots&\frac {k}{R}    
  \end{array}\right].
  \eeqa
  
  Following \cite{Ds} the diagonalizing of the mass matrix leads to the mixing of the kth state to the
  lowest mass eigenstate $\nu^m_{0L}$ given by
  \beq
  \label{eq:mixa}
     \tan {2\theta_k} \approx \frac{2\xi}{k-\xi^2} \;,
  \eeq
  for small $\xi$ 
where $\xi\equiv \sqrt{2}m_{D}R$. For values of $m_D\sim 10^{-4} \mathrm eV$ and 
$R\sim 1 \, {\rm mm}$ we 
  find $\xi\sim 10^{-1}$. Thus the mixing of the KK states with the lowest 
mass eigenstate $\nu_{0L}^m$ 
  becomes progressively smaller as the value of $k$ increases.  A solution of the solar
  neutrino problem in terms of matter enhanced flavor transformation of $\nu_{eL}$ 
  into these bulk modes has been explored in \cite{Ds}.  It is also studied
  in the supernova collapse phase in \cite{george}.  An alternate model of
neutrino mass and right handed neutrinos is provided in \cite{lor}.
  
  In this paper we study an important role that these bulk neutrino modes can play
  in stellar energy loss.   There are
several processes, such as the magnetic moment interaction, spin flip via the
ordinary weak interaction and Higgs boson exchange which can populate the bulk
neutrino modes, allowing energy to escape more quickly from stellar 
environments than in the case without neutrino mass.  
In the bulk neutrino scenario each mechanism is enhanced albeit 
  differently by the volume of the compactified space. This volume enhancement is due to 
the large 
  number of  KK states into which the active neutrino can scatter.
However, in Higgs exchange scattering, the large enhancement factor is 
insufficient to overcome the small  Higgs-fermion couplings.  In the following we
focus on the
 transition magnetic moment interaction for the active neutrinos and the
 bulk modes.  There will be a similar enhancement in the
case of neutrino spin flip scattering, but this process is more difficult to
study. In general, the rate of these processes is 
determined not only by the volume of the extra dimensions, 
but also by the energy available for the scattering.
  
  The same enhancement factor appears in the consideration of stellar energy loss due to KK graviton 
  states. This calculation involves theoretical modeling of KK graviton pion and nucleon couplings. 
  Model dependence in the case of bulk neutrinos enters through the
  coupling of the active neutrinos to the bulk modes and the choice of 
  Dirac or Majorana neutrinos.  Here for simplicity we assume only $\nu_{eL}$
  couples to the bulk modes, ignoring the couplings of the other two flavors.
  We have also taken the KK neutrinos to be Dirac neutrinos. 
  The limits we obtain below are complementary
  to the constraints from the graviton case although as we will show, 
  numerically they are similar.
  
  We begin by discussing the neutrino magnetic moment, $\mu_{(k)}$ generated by the 
  presence of the $k^{th}$ bulk neutrinos. This is given by the familiar magnetic 
  moment interaction term, 
  $ \mu_{(k)} \bar{\psi}_{\nu_{kR}^m}\sigma_{\lambda\rho}\psi_{\nu_{0L}^m} F^{\lambda\rho}$,
written here in terms of mass eigenstates. 
  The transition magnetic moment is calculated to be
  \beqa
  \label{eq:numag}
   \mu_{(k)}&=&\frac{\mathrm 3 \, e \, G_{F}}{\sqrt{2} (4 \pi)^2}\theta_{k}\left(\frac{k}{R}\right) \\
            &=& 1.6 \times 10^{-19} \mu_B \left( { m_D \over 1 {\rm eV}} \right)              \;
  \eeqa
  where we have used Eq. \ref {eq:mixa} for small $\xi$. It can be seen from this equation that the 
  magnetic moment is independent of $k$ and only depends on the value of $m_D$.  The same
result can be obtained by using the mass insertion technique and can therefore be
generalized to more than one dimension.

Electron neutrinos can transform to kth bulk neutrinos by way of the magnetic moment 
  interaction at a rate proportional to $ N \alpha \mu_{(k)}^2$ where N is the number of 
  available bulk neutrino modes, and $\alpha$ is the fine structure constant. 
  Transitions can take place as long as the mass of the mode is less than the 
  available energy, E.  In the case of an electron neutrino scattering into
  a bulk mode, the available energy is that of the incoming electron
  neutrino.  In the case of plasmon decay, the available energy is that of the
plasmon.  The total effective magnetic moment, $\mu_{eff}$ is a sum over 
  the full multiplicity of the available KK modes. Neglecting the influence of
  the mass of the neutrinos in the phase space factor,
 \beq
  \label{eq:munueff}
               \mu_{eff} \approx N^{1/2} \mu_{(k)}  
  \approx 4 \times 10^{-8}  \mu_B y\left( {10^{-5} \over 2 \pi} \right)^{n/2} A_n^{1/2} \left( {E \over 10 \, {\rm MeV}} \right)^{n/2} \left( {1 \, {\rm TeV} \over M_*} \right)^{n/2} \; .
  \eeq
The neutrino magnetic moment depends on both the energy of the neutrinos, 
  the number of extra dimensions and on the higher dimensional scale  $M_*$.  In equation
  \ref{eq:munueff}, $A_n$ is the volume of the positive hemisphere
of a n-dimensional unit sphere;  $A_n = 1$ for one
  extra dimension.  It can be seen 
  from the above equations that a larger $M_*$ or
  a larger number of  extra dimensions will
  make the effective neutrino magnetic moment smaller.
  A larger available energy, however, 
  will make the moment bigger, since there are many more 
  energetically allowed states for the bulk neutrinos.  

This effective magnetic moment expressed as Eq. \ref{eq:munueff} 
  {\em does not} depend on the manner in which the volume $V_n$ is distributed 
  among the extra dimensions.  It is also independent of the scaling relation 
Eq.~\ref{eq:fun}.  It is a consequence of the form of the Yukawa coupling given in 
Eq.~\ref{eq:hnu}.  If this coupling had a different dependence on volume, then
the effective magnetic moment could well depend on the scales of the
extra dimensions. In contrast, limits on the higher dimensional scale
from graviton emission are directly linked to Eq.~\ref{eq:fun}.  
  
  There are several existing limits on the the size of the neutrino magnetic moment coming
  from both terrestrial experiments and astrophysical considerations.  A review of the 
  astrophysical limits may be found in \cite{raffelt}.
  
  One such limit may be obtained from
  analysis of electron recoil spectra from neutrino-electron scattering 
  done with reactor antineutrinos.
  Reactor antineutrinos can have energies up to $\sim 10 \, {\rm MeV}$ but the peak in the spectrum
  comes at about $\sim 1 \, {\rm MeV}$.  At present the best limit on the neutrino
  magnetic moment from this method is $\mu_\nu = 1.9 \times 10^{-10} \mu_B$ 
  at $95 \%$ confidence \cite{reactor}.  Using equation \ref{eq:munueff} with
  $ E = 1 \, {\rm MeV}$, and a coupling of $y = 1$, we see that for $n=1$
  with a $M_{\ast}= 1$ TeV, $ \mu_{eff} \approx 2 \times 10^{-11} \mu_B$.  Such a 
  scenario is not ruled out by reactor neutrino experiments.  
  
  Similarly, a limit on the neutrino magnetic moment can be derived from the shape of the 
  electron recoil spectrum from neutrino - electron scattering by solar neutrinos 
  as measured at SuperKamiokande \cite{superk}.  
  The neutrinos which contribute to this scattering rate
  are about $5 - 15$ {\rm MeV}  and the limit on the neutrino magnetic moment is 
  $\mu_\nu \leq 1.6 \times 10^{-10} \mu_B$  \cite{beacom}.  Since these neutrinos
  are about an order of magnitude more energetic than the reactor neutrinos, for
  $M_* = 1 \, {\rm TeV}$, $y = 1$ and $n =1$, 
$ \mu_{eff} \approx 5 \times 10^{-11} \mu_B$.
  Therefore, these parameters are allowed by this measurement as well. 
  
  The effective neutrino magnetic moment for available energies of 
10  MeV is shown in Figure \ref{fig:mag_string}.  This
  figure  plots the neutrino magnetic moment against the scale $M_*$.  Effective magnetic 
  moments are shown for different numbers of extra dimensions.  Also shown as a horizontal line
  is the limit from the solar neutrino data.
  
  An astrophysical  limit on the  effective magnetic moment comes
  from plasmon decay \cite {Plas} in horizontal branch stars.  Plasmon decay
  through the  magnetic moment causes an increased energy loss in these stars and decreases
  their lifetime.  The limit on the neutrino magnetic moment from this phenomenon is   
  given by $\mu_{\nu}\leq 10^{-11} \mu_{\mathrm B}$ \cite {Fuku}.  The temperature of 
  the core of these stars is on the order of 10 keV.  Therefore around 10 keV is available for the
  neutrinos.  Using Eq \ref{eq:munueff} with E = 0.001, we see that the bulk neutrino
  effective magnetic moment contribution for plasmon decay will be considerably smaller than that
  for reactor and solar neutrinos.    
  
  On the other hand, supernova neutrinos could have energies up to 30 MeV 
in the core of the
  proto-neutron star.  If the neutrino magnetic moment is large, then 
  electron neutrinos can transform into right handed bulk neutrino states and
  escape from the core of the neutron star.  This releases energy from the proto neutron
  star too quickly  for the neutrino signature to be
  in agreement with the neutrino signal from supernova 1987a. The constraint on 
  the neutrino magnetic moment derived in this fashion 
  is $\mu_{\nu}\leq 10^{-12} \mu_{\mathrm B}$ \cite{BM}.  This limit is shown as 
a horizontal dashed line on Figure \ref{fig:mag_string}.
  
  The dependence of available energy on the effective magnetic moment is shown
  in Figure \ref{fig:mag_avail}.  The limits from supernova and HB stars are shown as crosses.
  The calculated effective magnetic moment is plotted as solid lines for the
  cases of one, two and three dimensions.  From Figures \ref{fig:mag_string} and 
  \ref{fig:mag_avail} it is clear
  that a single extra dimension, with a coupling of $y=1$ can not be
  reconciled with the supernova neutrino signal without further modification of the neutrino
  interaction and/or supernova picture.  Also apparent is the dramatic effect of
available energy on the effective magnetic moment.  Measurements of or
constraints on $\mu_\nu$ at high energy are more successful at probing 
this extra dimensional model.  We show explicitly the effect of $y \neq 1$ in 
Figure 3.   
  
  In conclusion we have shown that if $\nu_{\mathrm R}$ is a bulk neutrino it can 
induce an effective transition magnetic moment for the active neutrino which 
depends on the  number of extra dimensions.  
Neutrino magnetic moment considerations are a probe of the
coupling of the bulk neutrinos to the active neutrinos.  
The constraints discussed here 
 on extra dimensions can be considered 
  independently of the gravitational limits, since limits stemming from 
the neutrino magnetic moment do not depend on the gravitational 
scaling relation.  Bulk neutrino flavor transformation scenarios either in the
sun or in supernovae require at least two extra dimensions, a small Yukawa coupling
$y \simle 10^{-1}$ or a large higher dimensional scale $M_*$ in order
to avoid significantly altering the picture of 
supernova neutrino energy loss.

We thank the Institute for Nuclear Theory at the University of Washington
for its hospitality and the Department of Energy for partial support during the
completion of this work. This work is partially supported by the Natural Science 
and Engineering Research Council of Canada.

  \newpage
  
  \begin{figure}
  %\epsfxsize=14cm
  %\epsfxsize=14cm
  %\centerline{\epsffile{fig1.ps}}
  \caption{\label{fig:mag_string}
  Shows the effective neutrino magnetic moment in units on $\mu_B$, for the parameters
  $y=1$, $E=10 \, {\rm MeV}$.  The effective magnetic moment is plotted against $M_*$ 
  for n = 1, 2 and 3 dimensions.  Also shown as dashed lines are the limits on
  the moment from supernova neutrinos (lower line) 
  and neutrino electron scattering at SuperKamiokande (upper line).
  }
  \end{figure}
  
  \begin{figure}
  %\epsfxsize=14cm
  %\epsfxsize=14cm
  %\centerline{\epsffile{fig2.ps}}
  \caption{\label{fig:mag_avail}
  Shows the effective neutrino magnetic moment in units on $\mu_B$, for the parameters
  $y=1$, $M_*= 1 \, { \rm TeV}$.  The effective magnetic moment is plotted against available
  energy for n = 1, 2 and 3 dimensions.  Also shown as crosses are the limit
  the moment from supernova neutrinos (lower cross) 
  and lifetime of HB stars (upper cross).
  }
  \end{figure}

\begin{figure}
  %\epsfxsize=14cm
  %\epsfxsize=14cm
  %\centerline{\epsffile{fig2.ps}}
  \caption{\label{fig:mag_yukawa}
  Shows the effective neutrino magnetic moment in units on $\mu_B$, for the parameters
  $E =10 \, {\rm MeV}$, $M_*= 1 \, { \rm TeV}$.  The effective magnetic moment is plotted against 
Yukawa coupling for n = 1 and 2 dimensions. Also shown as horizontal lines are the limits on
  the moment from supernova neutrinos (lower line) 
  and neutrino electron scattering at SuperKamiokande (upper line).
  }
  \end{figure}
  
  \newpage
  
  \setcounter{figure}{0}
  \begin{figure}
  \epsfxsize=14cm
  \centerline{\epsffile{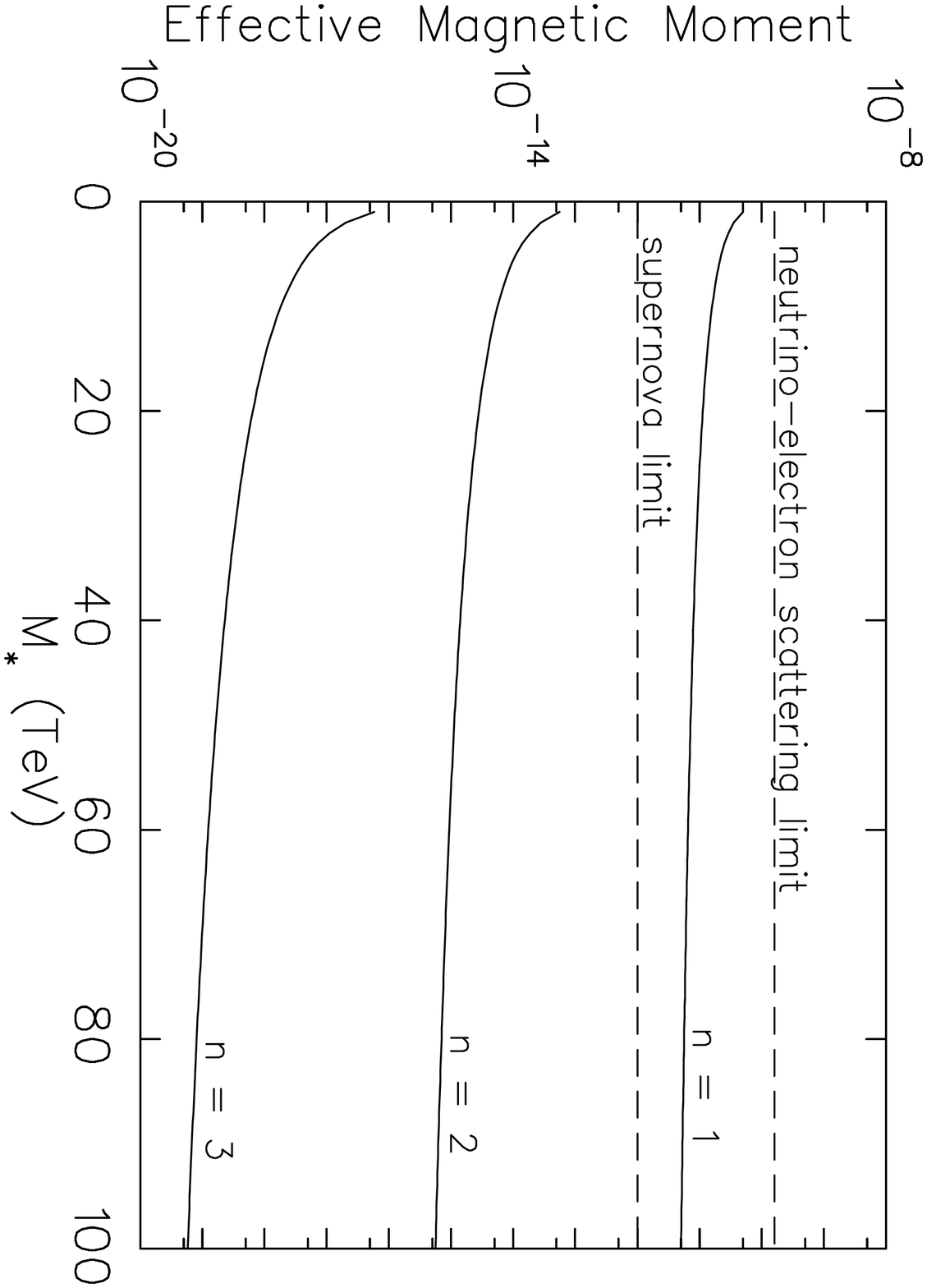}}
  %\caption{}
  \end{figure}
  
  \begin{figure}
  \epsfxsize=14cm
  \centerline{\epsffile{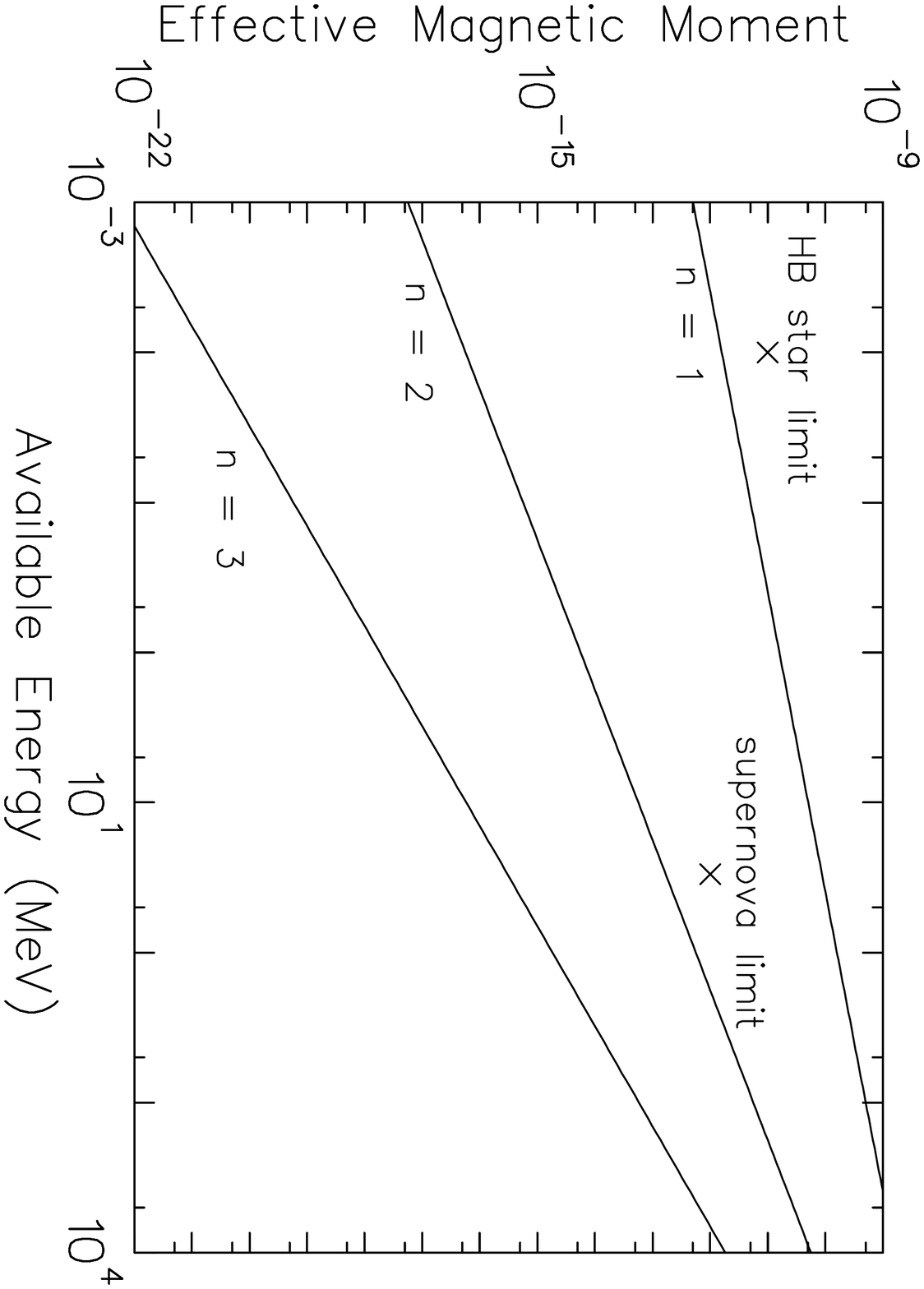}}
  %\caption{}
  \end{figure}

  \begin{figure}
  \epsfxsize=14cm
  \centerline{\epsffile{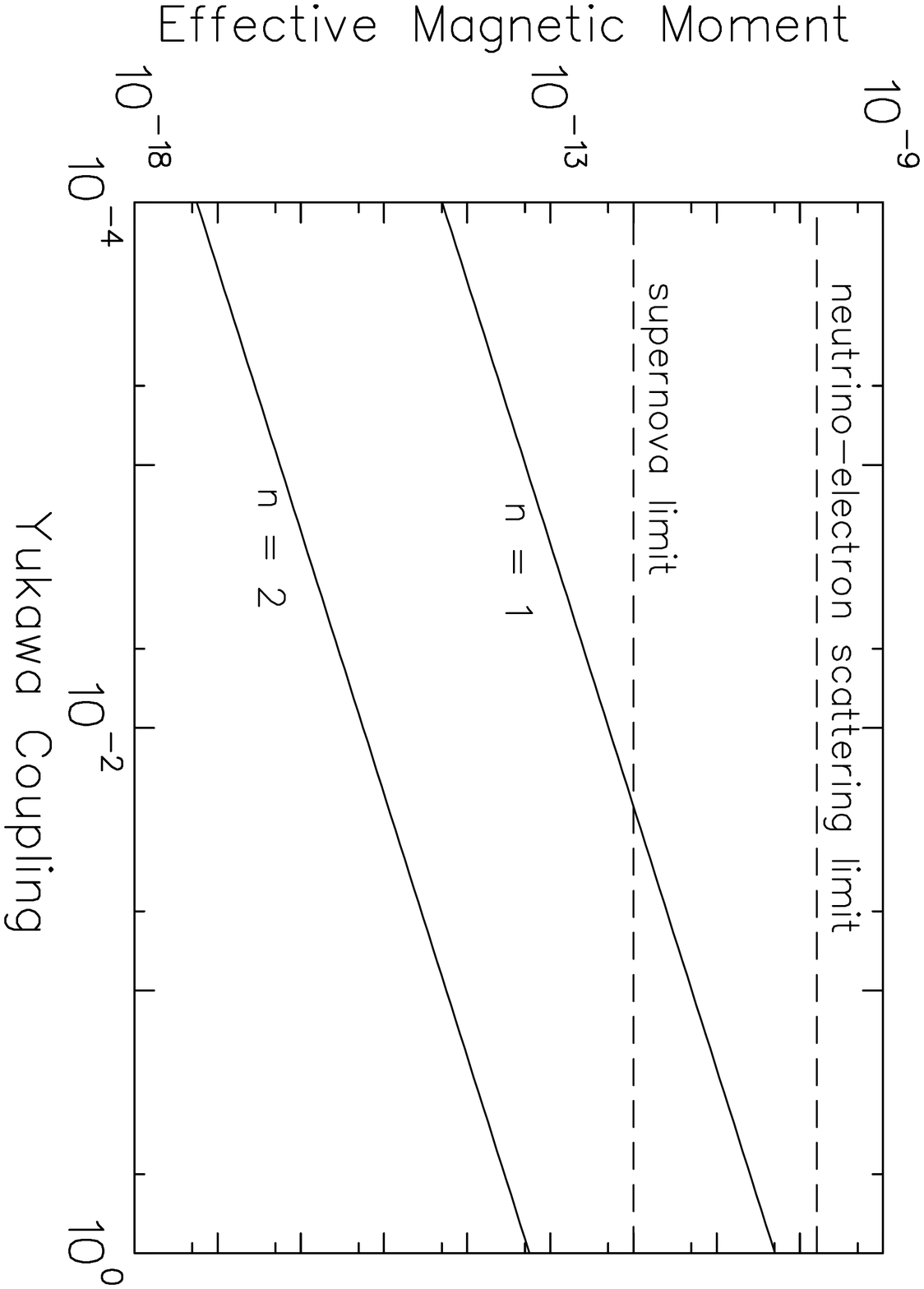}}
  %\caption{}
  \end{figure}


\begin{thebibliography}{99}
  \bibitem{Gin}P. Ginsparg, Phys. Lett B197 (1987) 139
  \bibitem{Wit}E. Witten, Nucl Phys. B471 (1996) 135
  \bibitem{Lyk}J. D. Lykken, Phys. Rev D54 (1996) 3693.
\bibitem{ant} I. Antoniadis, Phys. Lett. B246 (1990) 377 and 
I. Antoniadis, N. Arkani-Hamed, S. Dimopoulos and G. Dvali Phys. Lett. B436 (1998) 257.
  \bibitem{Nima}N. Arkani-Hamed, S. Dimopoulos, and G. Dvali, Phys. Lett. B429 (1998) 263, and
                Phys Rev. D59 (1999) 086004.
  \bibitem{Lon}J. C. Long, H.W. Chan, and J. C. Price , Nucl. Phys. B539 (1999) 23.
  \bibitem{Rand}L. Randall and R. Sundrum hep-ph/9905221
  \bibitem{Pol}J. Polchinski, hep-th/9611050
  \bibitem{Dva}G. Dvali and M. Shifman, Phys Lett B396(1997) 64.
  \bibitem{Cul}S. Cullen and M. Perelstein, Phys. Rev. Lett. 83 (1999) 268.
  \bibitem{many}K. Dienes, E. Dudas and T. Gherghetta, Phys. Lett B436 (1998) 55,\\
                G. Shiu and S-H. H. Tye, Phys Rev. D58 (1998)106007,\\
                Z. Kakushadze and S.H. Tye, hep-th/9806143 and hep-th/9809147,\\
                T. Banks, M. Dine, and A. E. Nelson, hep-th/9903019, and\\
                Ref 4.
  \bibitem{Supk}Y. Fukuda et al., Phys. Rev Lett. 82 (1998) 2644.\\
                For more recent data see K. Scholberg, hepph 9905016.
  \bibitem{Nima2}N. Arkani-Hamed, S. Dimopoulos, G. Dvali and J. March-Russel, hep-ph/9811448.
  \bibitem{Die}K. R. Dienes, E. Dudas, and T. Gherghetta, hep-ph/9811428.
\bibitem{pos} A. Faraggi and M. Pospelov Phys.Lett. B458 237-244, 1999.
  \bibitem{Ds}G. Dvali and A.Yu. Smirnov hep-ph/9904211. 
   \bibitem{george} G. M. Fuller and M. Patel, in preparation.
\bibitem{lor} R. N. Mohapatra, S. Nandi and  A. Prez-Lorenzana, hep-ph/9907520.
  \bibitem{raffelt} G. G. Raffelt, Stars as Laboratories for Fundamental Physics 
  (University of Chicago Press, Chicago 1996).
  \bibitem{reactor} A. I. Derbin et al., JETP Lett 57 (1993) 768.
  \bibitem{superk} M. B. Smy. Proceedings for DPF'99 Conference (1999).
  \bibitem{beacom} J. F. Beacom and P. Vogel, hep-ph/9907383.
  \bibitem{Plas}P. Sutherland, J.N. Ng, E. Flowers, M. Ruderman, and C. Inman, 
Phys. Rev. D13 (1976) 2700.  
  \bibitem{Fuku} M. Fukugita and S. Yazaki, Phys. Rev. D36 (1987) 3817.
  \bibitem{BM}R. Barbieri and R. N. Mohapatra, Phys. Rev. Lett. 61 (1988) 27, \\
              J. M. Lattimer and J. Cooperstein, Phys. Rev. Lett. 61 (1988) 23.
  
  \end{thebibliography}
  \end{document}